 \documentclass[prd,aps,
twocolumn,
nofootinbib,
floatfix,
superscriptaddress]{revtex4}
\usepackage{graphicx}
\usepackage{epsfig}
\usepackage{rotating}
\usepackage{amssymb}
\usepackage{subfigure}
\usepackage{dsfont}
\usepackage{psfrag}
\usepackage{amsmath,euscript,array,mathrsfs}
\usepackage{xcolor}

\newcommand{\beq}{\begin{equation}}
\newcommand{\eeq}{\end{equation}}
\newcommand{\beqs}{\begin{eqnarray}}
\newcommand{\eeqs}{\end{eqnarray}}

\newcommand{\gsim}{\mathrel{\raisebox{-
.6ex}{$\stackrel{\textstyle>}{\sim}$}}}

\def\hbar{\hspace{0pt}\raisebox{1pt}{$-$} \hspace{-7pt} h}

\def\di{\mbox{d}}
\def\r{\rho}

\newcommand{\be}{\begin{equation}}
\newcommand{\ee}{\end{equation}}

\newcommand{\bea}{\begin{eqnarray}}
\newcommand{\eea}{\end{eqnarray}}

\def\lbldef#1#2{\expandafter\gdef\csname #1\endcsname {#2}}

\def\href#1#2{#2}

\newcommand{\ber}{\begin{eqnarray}}
\newcommand{\eer}{\end{eqnarray}}

\newcommand{\beqar}{\begin{eqnarray}}

\newcommand{\eeqar}{\end{eqnarray}}

\newcommand{\dsl}
  {\kern.06em\hbox{\raise.15ex\hbox{$/$}\kern-.56em\hbox{$\partial$}}}

\newcommand{\eeqarr}{\end{eqnarray}}
\newcommand{\ZZ}{{\rm \kern 0.275em Z \kern -0.92em Z}\;}

\def\CC{{\mathchoice
{\rm C\mkern-8mu\vrule height1.45ex depth-.05ex
width.05em\mkern9mu\kern-.05em}
{\rm C\mkern-8mu\vrule height1.45ex depth-.05ex
width.05em\mkern9mu\kern-.05em}
{\rm C\mkern-8mu\vrule height1ex depth-.07ex
width.035em\mkern9mu\kern-.035em}
{\rm C\mkern-8mu\vrule height.65ex depth-.1ex
width.025em\mkern8mu\kern-.025em}}}

\def\RR{{\rm I\kern-1.6pt {\rm R}}}

\def\ZZ{{\rm Z}\kern-3.8pt {\rm Z} \kern2pt}
\def\IB{\relax{\rm I\kern-.18em B}}
\def\ID{\relax{\rm I\kern-.18em D}}
\def\II{\relax{\rm I\kern-.18em I}}
\def\IP{\relax{\rm I\kern-.18em P}}

\newcommand{\bear}{\begin{eqnarray}}
\newcommand{\eear}{\end{eqnarray}}

\def\to{\rightarrow}

\def\to{\rightarrow}

\def\c{\gamma}

\def\f{\phi}               

\def\r{\rho}                                     

\def\6{\partial}

\def\bea{\begin{eqnarray}}
\def\eea{\end{eqnarray}}

\def\beqx{\begin{displaymath}}
\def\eeqx{\end{displaymath}}

\newcommand{\bmat}{\left(\begin{array}}
\newcommand{\emat}{\end{array}\right)}

\def\c{\chi}

\def\f{\phi}

\def\r{\rho}

\def\bo{{\raise-.3ex\hbox{\large$\Box$}}}               
\def\face{{\raise.2ex\hbox{$\displaystyle \bigodot$}\mskip-2.2mu \llap {$\ddot
        \smile$}}}                                   
\def\>{\rangle}                                      
\def\<{\langle}                                      

\def\leftrightarrowfill{$\mathsurround=0pt \mathord\leftarrow \mkern-6mu
        \cleaders\hbox{$\mkern-2mu \mathord- \mkern-2mu$}\hfill
        \mkern-6mu \mathord\rightarrow$}        
\def\dvec#1{\vbox{\ialign{##\crcr
        \leftrightarrowfill\crcr\noalign{\kern-1pt\nointerlineskip}
        $\hfil\displaystyle{#1}\hfil$\crcr}}}           

\def\-{\hphantom{-}}

\begin{document}

\title{A light dilaton in a metastable vacuum}

\author{Daniel Elander}
\affiliation{Laboratoire Charles Coulomb (L2C), University of Montpellier, CNRS, Montpellier, France}

\author{Maurizio Piai}
\affiliation{Department of Physics, College of Science, Swansea University,
Singleton Park, SA2 8PP, Swansea, Wales, UK}

\author{John Roughley}
\affiliation{Department of Physics, College of Science, Swansea University,
Singleton Park, SA2 8PP, Swansea, Wales, UK}

\date{\today}

\allowdisplaybreaks


\begin{abstract}

We identify a parametrically light dilaton by studying the perturbations 
of metastable vacua along 
 a branch of regular supergravity backgrounds 
 that are dual to four-dimensional confining field theories.
The branch includes 
also stable  and 
 unstable solutions. 
 The former encompass, as a special case, the geometry proposed by 
 Witten as a holographic model of confinement. 
  The latter approach a supersymmetric solution, by
  enhancing a condensate in the dual field theory.
  A phase transition separates the space of stable 
 backgrounds from the metastable ones. 
 In proximity of the phase transition, one of the lightest scalar states inherits
  some of the properties of the dilaton, despite not being particularly light.

\end{abstract}

\maketitle

\section{Introduction}

The Higgs particle~\cite{Aad:2012tfa,Chatrchyan:2012xdj} might originate as a composite 
dilaton in a new strongly coupled  theory. The literature on the effective field theory description of the dilaton
has an  ancient origin~\cite{Migdal:1982jp,Coleman:1985rnk}. It has been invoked 
in the context of dynamical electroweak symmetry
 breaking~\cite{Leung:1985sn,Bardeen:1985sm,Yamawaki:1985zg}, of extensions of the standard 
model~\cite{Goldberger:2008zz,Hong:2004td,Dietrich:2005jn,Hashimoto:2010nw,
Appelquist:2010gy,Vecchi:2010gj,Chacko:2012sy,
 Bellazzini:2012vz,Bellazzini:2013fga,Abe:2012eu,Eichten:2012qb,Hernandez-Leon:2017kea},
and in the interpretation of lattice data~\cite{Matsuzaki:2013eva,
Golterman:2016lsd,Kasai:2016ifi,Hansen:2016fri,Golterman:2016cdd,Appelquist:2017wcg,Appelquist:2017vyy,
Golterman:2018mfm,Cata:2019edh,Appelquist:2019lgk,Cata:2018wzl}.
With the advent of gauge-gravity dualities~\cite{Maldacena:1997re,Gubser:1998bc,Witten:1998qj,Aharony:1999ti},
holographic models giving rise to a dilatonic state have been identified and studied both in the 
context of bottom-up~\cite{Goldberger:1999uk,DeWolfe:1999cp,Goldberger:1999un,Csaki:2000zn,
ArkaniHamed:2000ds,Rattazzi:2000hs,Kofman:2004tk,Elander:2011aa,Kutasov:2012uq,Lawrance:2012cg,
Elander:2012fk,Goykhman:2012az,Evans:2013vca,Megias:2014iwa,Elander:2015asa}
and top-down constructions derived from 
supergravity~\cite{Elander:2009pk,Elander:2012yh,Elander:2014ola,Elander:2017cle,Elander:2017hyr}.

We  pursue an alternative approach to the study of the dilaton,
along the programme announced in Ref.~\cite{Elander:2020ial},
which is inspired by  Refs.~\cite{Kaplan:2009kr,Gorbenko:2018ncu,
Gorbenko:2018dtm,Pomarol:2019aae}, but is implemented
within the rigorous framework of supergravity. 
We generalise the notion of proximity to the
BF unitarity bound~\cite{Breitenlohner:1982jf}---central to the arguments in 
Ref.~\cite{Kaplan:2009kr}---in order  to explore non-AdS backgrounds dual 
to confining theories, in regions of parameter space near
tachyonic instabilities. We aim at ascertaining whether the
 spectrum of bound states includes a light dilaton.

In this paper we consider the toroidal compactifiction of the maximal 
supergravity theory in $D=7$ dimensions~\cite{Nastase:1999cb,
Pernici:1984xx,Pernici:1984zw,Lu:1999bc,Campos:2000yu}, that admits as 
a background solution the holographic description of 
confinement proposed by Witten~\cite{Witten:1998zw}---also used for phenomenological purposes
by Sakai and Sugimoto~\cite{Sakai:2004cn,Sakai:2005yt}.
We compare to the case of Romans theory~\cite{Romans:1985tw}---see
Refs.~\cite{Elander:2020ial,Elander:2013jqa,Elander:2018aub}.

We focus on three  branches of solutions: 
i) regular solutions that include the Witten model and are
 interpreted as duals of four-dimensional confining theories, 
ii) a class of supersymmetric solutions, and iii) a branch of non-supersymmetric solutions,
 that  (locally) preserve six-dimensional Poincar\'e invariance,
but are badly singular---they do not even meet Gubser's criteria~\cite{Gubser:2000nd}.
We compute the spectrum of fluctuations of the 
relevant scalar and spin-2 tensor states, using the gauge-invariant formalism of Refs.~\cite{Bianchi:2003ug,
Berg:2005pd,Berg:2006xy,Elander:2009bm,Elander:2010wd}, hence extending the study of the spectra
performed in Ref.~\cite{Brower:2000rp} and Ref.~\cite{Elander:2013jqa}.
We compare to the result of applying the probe approximation~\cite{Elander:2020csd}, in order to ascertain
 whether any of the scalar states have significant overlap with the trace of the stress-energy tensor, 
 and can hence be identified with an approximate dilaton.
 
In a  region of parameter space  the spectrum contains 
  a parametrically light dilaton. We study  the energetics along the three branches of solutions,
by computing the free energy using  holographic renormalisation~\cite{Bianchi:2001kw,
Skenderis:2002wp,Papadimitriou:2004ap}, and employing a simple scale-setting procedure to
 compare different backgrounds~\cite{Csaki:2000cx}. We present firm evidence of the existence of a phase transition
 in the gravity theory (see also Ref.~\cite{Faedo:2014naa}).
The parametrically light dilaton emerges along the portion of the regular branch of solutions
which contains metastable solutions,
the life-time of which is not known (but see Ref.~\cite{Bigazzi:2020phm}).

\section{The Gravity Model}
\label{Sec:Action}

\begin{figure}[t]
	\centering
	\begin{picture}(300,153)
	\put(-5,0){\includegraphics[width=9cm]{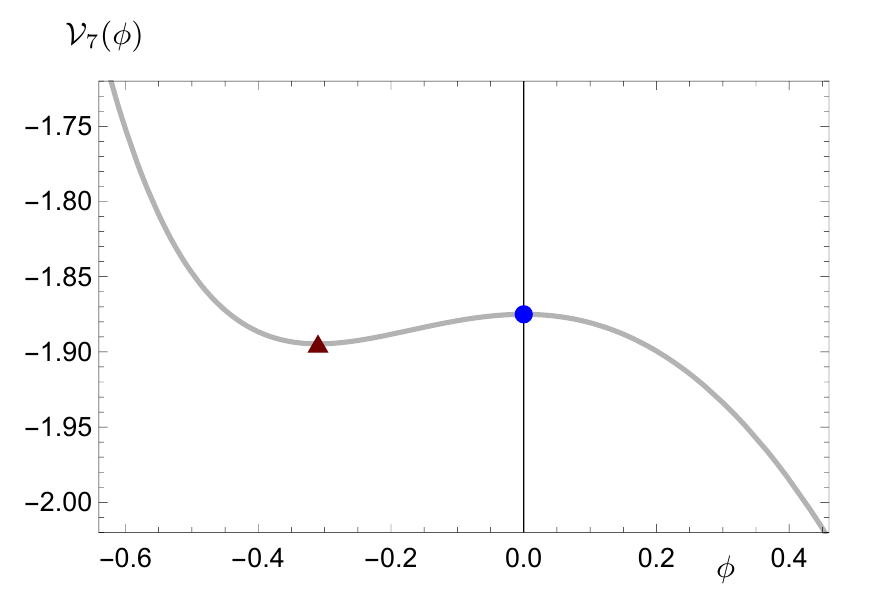}}
	\end{picture}
	\caption{The potential ${\cal V}_7(\phi)$ of the theory. The blue disk is the $\phi_{UV}$ critical point,
	while the red triangle is the $\phi_{IR}$ one.}
	\label{Fig:potential}
\end{figure}

We denote with hatted symbols quantities characterising the theory in $D=7$ dimensions.
The action, 
truncated to retain the scalar $\phi$
coupled to gravity, is the following~\cite{Elander:2020csd} (see also
Refs.~\cite{Elander:2013jqa,Pernici:1984xx,Pernici:1984zw}):
\beqs
{\cal S}_7=\int\di^7x \sqrt{-\hat{g}_7}\left[\frac{{\cal R}_7}{4}-\frac{\hat{g}^{\hat{M}\hat{N}}}{4}\partial_{\hat{M}}
\phi\partial_{\hat{N}}\phi -{\cal V}_7\right],
\label{Eq:Action7D}
\eeqs
where
 the potential (see Fig.~\ref{Fig:potential}) is
\beqs
{\cal V}_7&=&\frac{1}{8}e^{-\frac{8}{\sqrt{5}} \phi} - e^{-\frac{3}{\sqrt{5}}\phi} -e^{\frac{2}{\sqrt{5}}\phi}\,.
\eeqs
This potential admits two critical  points. The one with $\phi=\phi_{UV}=0$ will play a central role in this paper,
as it corresponds to a UV fixed point in the dual field theory. It yields ${\cal V}_7(\phi_{UV})=-\frac{15}{8}$.
Another  critical point of ${\cal V}_7$  has
$\phi_{IR}=-\frac{1}{\sqrt{5}}\log (2)$, for which ${\cal V}_7(\phi_{IR})=-\frac{5}{2^{7/5}}$.

Following the notation in Refs.~\cite{Elander:2020csd,Elander:2013jqa},
we reduce to $D=5$ dimensions  by adopting the following ansatz:
\beqs
\di s_7^2&=&e^{-2\chi}\di s_5^2 \,+\,e^{3\chi-2\omega}\di \eta^2 \,+\, e^{3\chi+2\omega} \di \zeta^2\,,
\eeqs
where the  metric $\di s_5^2$ takes the domain wall form
\beqs
\di s_5^2 &=&e^{2A(r)}\di x_{1,3}^2 \,+\,\di r^2\,,
\eeqs
and the background profiles $\phi(r)$, $\chi(r)$, $\omega(r)$, and $A(r)$
depend only on the the radial coordinate $r$.  The angles $0\leq \eta,\zeta<2\pi$ parametrise a torus.
We  apply the change of variables $\di \r =e^{-\chi} \di r$.
The domain-wall (DW) ansatz in $D=7$ dimensions is recovered
by  imposing the  constraints
$\omega\,=\,0$ and 
${\cal A}=A-\chi=\frac{3}{5} A=\frac{3}{2}\chi$,
and hence the AdS$_7$ solution has $\partial_{\rho} {\cal A}=\frac{1}{2}$, 
$\partial_{\rho}\chi=\frac{1}{3}$, and $\partial_{\rho}A=\frac{5}{6}$.
The bulk action in $D=5$ dimensions  is the following:
\beqs
{\cal S}_5=
\int\di^5x \sqrt{-{g}_5}\left[\frac{{\cal R}_5}{4}-\frac{{g}^{{M}{N}}}{2}{G}_{ab}\partial_{{M}}
{\Phi}^a\partial_{{N}}{\Phi}^b -{\cal V}\right],
\eeqs
where ${\cal V}=e^{-2\chi} {\cal V}_7$, $\Phi^a=\{\phi,\omega,\chi\}$,
and the sigma-model metric is $G_{ab}={\rm diag}\,\left(\frac{1}{2},1,\frac{15}{4}\right)$.
We verified that 
\beqs
{\cal S}_7=\int\di \eta\di \zeta\left[ {\cal S}_5+\int\di^5 x \partial_M \left(\frac{1}{2}\sqrt{-g_5}g^{MN}\partial_N\chi\right)\right]\,.
\eeqs

\section{Classes of Solutions}

All the solutions of interest approach $\phi=\phi_{UV}=0$  at large $\rho$.
We write them as a power 
series of the small coordinate $z\equiv e^{-\r/2}$, as follows
\begin{widetext}
\beqs
\phi(z)&=& \label{Eq:UV1}
\phi_2 z^2+ \left(\phi_4-\frac{18 \phi_2^2 \log (z)}{\sqrt{5}}\right)z^4+
   \left(\frac{162}{5} \phi_2^3 \log (z)-\frac{637 \phi_2^3}{30}-\frac{9 \phi_2
   \phi_4}{\sqrt{5}}\right)z^6+\cdots\,,
\\
   \omega(z)&= &\label{Eq:UV2}
\omega_{U}+\omega_{6}z^6+\cdots
\,,\\
\chi(z)&=&
 \chi_{U}-\frac{2}{3}\log (z)-\frac{\phi_2^2 z^4}{30}\,
+\frac{2  }{675}
  \left(\frac{675}{2}\chi_6-150\,\omega_{6}+72 \sqrt{5} \phi_2^3 \log (z)-6 \sqrt{5} \phi_2^3-20
   \phi_2 \phi_4\right)z^6+\cdots
   \label{Eq:UV3}
\,,\\
A(z)&=&
A_{U}-\frac{5}{3}\log (z)-\frac{\phi_2^2
   z^4}{12}+\frac{ 1}{270} \left(\frac{135}{2}\chi_6-30\,\omega_{6}+144 \sqrt{5} \phi_2^3 \log (z)-12
   \sqrt{5} \phi_2^3-40 \phi_2 \phi_4\right)z^6+\cdots
   \label{Eq:UV4}\,.
\eeqs
\end{widetext}

They are characterised by seven integration constants: $\phi_2$, $\phi_4$, $\omega_U$, $\omega_6$, $\chi_U$,
$\chi_6$, and $A_U$. 
The  DW solutions  have $\omega_U=\omega_6=\chi_6=0$
 and $\chi_U=\frac{2}{5}A_U$, leaving $A_U$, $\phi_2$ and $\phi_4$ as independent non-trivial free parameters.
What we will call confining solutions have $\chi_6=0$.

We find it convenient to define a scale $\Lambda$ as follows~\cite{Csaki:2000cx}:
\begin{equation}
\label{Eq:Scale}
\Lambda^{-1}\equiv
\int_{\rho_o}^{\infty}\di \rho\,e^{\chi(\rho)-A(\r)}\,,
\end{equation}
with $\r_o$ the end of space. While other choices might be admissible, this
has the advantage of being applicable to all the solutions of interest.

\subsection{SUSY solutions}

The supersymmetric DW solutions satisfy the following first-order differential equations:
\beqs
\partial_{\rho}{\cal A}=-\frac{2}{5}{\cal W}_1\,,\,\,\,\,\,
\partial_{\rho}\phi=2\partial_{\phi} {\cal W}_1\,.
\eeqs
The superpotential ${\cal W}_1\equiv-\frac{1}{4}e^{-\frac{4}{\sqrt{5}}\phi}\,-\,e^{\frac{1}{\sqrt{5}}\phi}$
 solves the defining equation
${\cal V}_D=\frac{1}{2}G^{\phi\phi}(\partial_{\phi}{\cal W})^2\,-\,\frac{D-1}{D-2}{\cal W}^2$ for $D=7$.
After performing the change 
of variables $\partial_{\rho}\equiv e^{-\frac{3\phi}{2\sqrt{5}}}\partial_{\tau}$,
we find the exact solutions 
\beq
\phi(\tau)=\frac{4}{\sqrt{5}}{\rm arctanh}\left(\frac{}{}e^{-2(\tau-\tau_o)}\right)\,,
\eeq
with the warp factor given by
\beq
{\cal A}(\tau)={\cal A}_o+\frac{1}{10}\log [\cosh(\tau-\tau_o)\sinh^4(\tau-\tau_o) ]\,,
\eeq
 where ${\cal A}_o$ and $\tau_o$ are real integration constants.

The IR expansion of these solutions in terms of the radial coordinate $\r$
and the new constants $\r_o$ and ${\cal A}_I={\cal A}_0+\log(\frac{2}{5})$ can be written explicitly
in the following form
\beq
\phi(\r)=
-\sqrt{5}\log\frac{2(\r-\r_o)}{5}+\frac{16\,(\r-\r_o)^5}{1875\sqrt{5}}+\cdots\,,
\eeq
and
\beq
{\cal A}(\r)=
{\cal A}_I+\log(\r-\r_o)+\frac{8\,(\r-\r_o)^5}{9375}+\cdots\,.
\eeq
Their holographic interpretation involves
an operator of dimension $\Delta=4$ developing a vacuum expectation value (VEV)
in the dual field theory.

The conjugate superpotential entering the calculation of the free energy
is known as a perturbative expansion:
\beqs
{\cal W}_2&=&-\frac{5}{4}-\frac{\phi^2}{4}+\frac{3 \phi^3}{4 \sqrt{5}} \log \left(\frac{\phi^2}{\kappa} \right)
   +\cdots\,,\label{Eq:W2}
\eeqs
where $\kappa$ is scheme dependent.

\subsection{Singular DW solutions}

A class of singular DW solutions is characterised by the harmless ${\cal A}_I$,
the end of space $\r_o$, and the non-trivial $\phi_5$. 
As anticipated, these solutions are badly singular: their Ricci scalar tensor ${\cal R}_7$
diverges, and the potential is not bounded from above, violating the requirement from Ref.~\cite{Gubser:2000nd}.
The IR expansion of solutions of this class reads as follows:
\begin{widetext}
\beqs
\phi(\rho)
&=&
\frac{ \sqrt{5}}{4}  \log\frac{8}{5} (\r-\r_o)
+\phi_5(\r-\r_o)^{5/8}+\frac{1}{135} \left(37 \sqrt{5}   \phi_5^2-24\ 2^{3/4} \sqrt[4]{5}\right)(\r-\r_o)^{5/4}+\cdots
\,,\\
{\cal A}(\rho)
&=&
{\cal A}_{I}
+\frac{\log (\r-\r_o)}{16}
+\frac{4 \phi_5}{3\sqrt{5}}(\r-\r_o)^{5/8}
+\frac{1}{2700}\left(192\times 10^{3/4}-2155 \phi_5^2\right)(\r-\r_o)^{5/4}+\cdots
\,.
\eeqs
\end{widetext}

\subsection{Confining  solutions}
The regular solutions of this class obey the constraint $A=\frac{5}{2}\chi+\omega$.
They depend on two harmless constants $\chi_I$ and $\omega_I$, besides
 $\r_o$ and $\phi_I$.
The IR expansion of these solutions reads as follows:
\begin{widetext}
\beqs
\phi(\r)&=&
\phi_I
   -\frac{1}{2 \sqrt{5}}(\r-\r_o)^2
   e^{-\frac{8 \phi_I}{\sqrt{5}}} \left(-3 e^{\sqrt{5} \phi_I}+2 e^{2
   \sqrt{5} \phi_I}+1\right)+\cdots
\,,\\
\omega(\r)&=&
\omega_I-\frac{\log
   (\r-\r_o)}{2}
   +\frac{1}{40} (\r-\r_o)^2 e^{-\frac{8 \phi_I}{\sqrt{5}}} \left(8
   e^{\sqrt{5} \phi_I}+8 e^{2 \sqrt{5} \phi_I}-1\right) +\cdots
   \,,\\
\chi(\r)&=&
\chi_I   +\frac{\log (\r-\r_o)}{3}
  +\frac{1}{6000}(\r-\r_o)^4 e^{-\frac{16 \phi_I}{\sqrt{5}}} 
 \left(32 e^{\sqrt{5} \phi_I}-56 e^{2 \sqrt{5} \phi_I}+224 e^{3 \sqrt{5} \phi_I}+32 e^{4 \sqrt{5} \phi_I}-7\right)+\cdots
\,.
\eeqs
\end{widetext}

We restrict attention to solutions flowing from the UV critical point, which requires
$\phi_I>\phi_{IR}$.
The invariants ${\cal R}_7$,  ${\cal R}_{7\,\hat{M}\hat{N}}{\cal R}_{7}^{\hat{M}\hat{N}}$, and 
${\cal R}^{\,\,\,\,\,\hat{P}}_{7\,\,\,\,\,\hat{M}\hat{N}\hat{Q}}{\cal R}_{7\,\hat{P}}^{\,\,\,\,\,\,\,\,\,\hat{M}\hat{N}\hat{Q}}$ are 
 finite.
We  impose the constraint $\omega_I=\frac{3}{2}\chi_I$ in order to avoid a conical singularity.

\section{Glueball masses}

We compute the spectrum of fluctuations of the five-dimensional theory, by employing the  gauge-invariant formalism 
developed in Refs.~\cite{Bianchi:2003ug,Berg:2005pd,Berg:2006xy,Elander:2009bm,Elander:2010wd}. 
We introduce the IR regulator  $\r_{1}$ with $\r_{o}<\r_{1}$, and  the UV regulator $\r_{2}$. 
The physical results are recovered in the limits $\r_1 \rightarrow \r_o$ 
and $\r_2 \rightarrow +\infty$ (see Refs.~\cite{Elander:2010wd,Elander:2013jqa,Elander:2018aub}). The  scalar fluctuations 
are written as the gauge invariant combinations
\beqs
\mathfrak{a}^{a}(M,\rho)&\equiv&\varphi^{a}(M,\rho)-\frac{\partial_{\rho}{\Phi}^{a}(\rho)}{6\partial_{\rho}A(\rho)}h(M,\rho)\,,\\
\label{Eq:inv}
\nonumber
\eeqs
where $M$ is the mass in the dual theory,
$\varphi^a$ are fluctuations  of the scalars $\Phi^a$ 
and $h$ of the trace of the four-dimensional portion of the metric. 
They obey the following linearised equations and boundary conditions:
\begin{widetext}
\beqs
\label{Eq:a}
0&=&\left[\frac{}{}e^{\chi}\partial_{\rho}(e^{-\chi}\partial_{\rho}) + (4\partial_{\rho}A)\partial_{\rho}
+e^{2\chi-2A}M^{2}\right]\mathfrak{a}^{a} - e^{2\chi} \mathcal X^{a}_{\ c}
\mathfrak{a}^{c}\,,\\
	\label{Eq:bc}
	0
	&=&
	\frac{}{}e^{-2\chi}\partial_{\rho} \Phi^{c}
	\partial_{\rho} \Phi^{d}G_{db}\partial_{\rho}
	\mathfrak{a}^{b}+\left.\left[\frac{}{}\frac{3\partial_{\rho}A}{2}e^{-2A}M^{2}\delta^{c}_{\ b}
	-\partial_{\rho} \Phi^{c}\bigg(\frac{4\mathcal{V}}{3\partial_{\rho}A}\partial_{\rho} \Phi^{d}G_{db} + 
	\frac{\partial\mathcal{V}}{\partial\Phi^{b}} \bigg) \right]\mathfrak{a}^{b}\right|_{\rho_{i}}\,,
\eeqs
\end{widetext}
where in all these expressions 
the quantities $A$, ${\Phi}^a$, and ${\cal V}$ are evaluated on the background, and  
\begin{widetext}
\beqs
\label{Eq:aa}
\mathcal X^{a}_{\ c}&\equiv&
 \partial_{c}\left[G^{ab}\frac{\partial \mathcal{V}}{\partial  \Phi^{b}}\right]\, 
+\frac{4}{3\partial_{\rho} A}
\bigg[\partial_{\rho} \Phi^{a}\frac{\partial \mathcal{V}}{\partial  \Phi^{c}}
+G^{ab}\frac{\partial \mathcal{V}}{\partial  \Phi^{b}}\partial_{\rho}\Phi^{d}G_{dc}\bigg]
+\frac{16\mathcal{V}}{9(\partial_{\rho}A)^{2}}\partial_{\rho} \Phi^{a}\partial_{\rho} \Phi^{b}G_{bc}\,.
\eeqs
\end{widetext}

The gauge invariant spin-2 tensor fluctuations obey the linearised equation
\beqs
0&=\left[\frac{}{}\partial^{2}_{\r}+(4\partial_{\r}A-\partial_{\r}\c)\partial_{\r}+e^{2\c-2A}M^{2}\right]\mathfrak{e}^{\mu}_{\ \nu}\,,
\label{Eq:e}
\eeqs
and Neumann boundary conditions $\left.\partial_\rho\mathfrak{e}^{\mu}_{\ \nu}\right|_{\rho_i}=0$.

The probe approximation for the scalars is defined by ignoring the term proportional to $h$ in Eq.~(\ref{Eq:inv}). 
According to the dictionary of gauge-gravity dualities, $h$ is the bulk field associated with the trace of the stress-energy tensor,
which is the field theory operator associated with dilatation, and sourcing the dilaton. Hence, this approximation holds for
 scalar bound states that decouple from the dilatation operator, and cannot be interpreted as a dilaton.
The equations for the scalar fluctuations greatly simplify, as only the first term in Eq.~(\ref{Eq:aa}) survives, 
and the boundary conditions reduce to Dirichlet. Note that the probe approximation is used solely as a diagnostic tool to identify scalar states which mix non-trivially with the dilaton.

\begin{figure}[b]
\begin{center}
	\begin{picture}(300,170)
	\put(-5,0){\includegraphics[width=9cm]{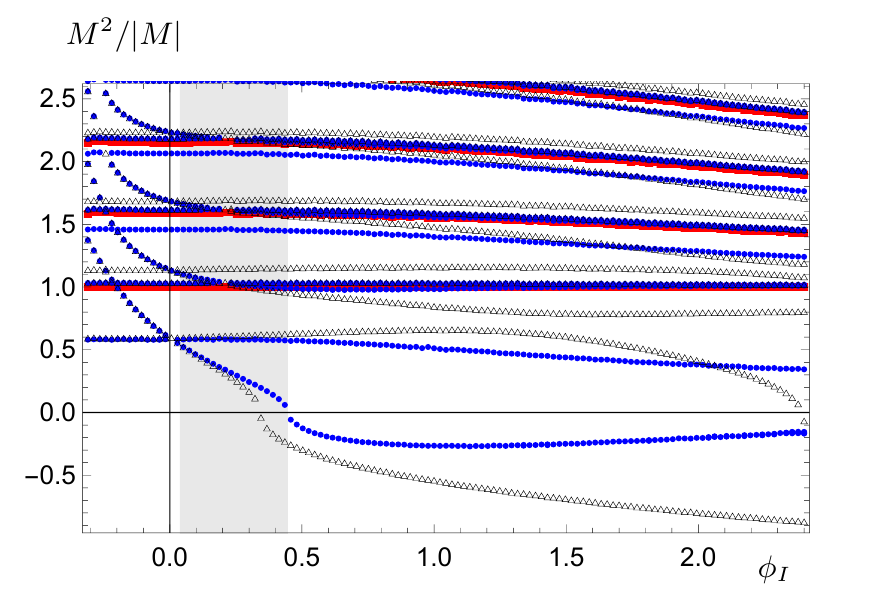}}
	\end{picture}
\caption{The spectra of scalar masses $M$ as a function of the parameter $\phi_{I}$
	along the confining branch of solutions, normalised in units of the lightest tensor mass, and using regulators $\r_{1}=10^{-4}$ and $\r_{2}=12$. 
	The (red) squares represent the spin-2 tensor modes, 
	the (blue) disks are the scalar, gauge invariant fluctuations originating from $\phi$,
	 $\chi$, and $\omega$. The (black) triangles do not represent an additional set of states: they
	 denote the same scalars, but computed in the probe approximation---neglecting the fluctuation of the 
	 background metric.
	The shading denotes the stable (leftmost white), metastable (grey), and unstable (rightmost white) backgrounds. We verified that our choices for the two regulators were sufficiently close to the physical limits ($\r_{1}\to\r_{o}$, $\r_{2}\to\infty$) to avoid discernible cutoff effects.}
\label{Fig:SpectrumProbe}
\end{center}
\end{figure}

In Fig.~\ref{Fig:SpectrumProbe}, we show the spectra of tensors and scalars, compared to the probe approximation,
normalised to the lightest spin-2 fluctuation. For $\phi_I<0$ the scalars agree with Ref.~\cite{Elander:2013jqa}. The new 
results for $\phi_I>0$ show that one of the scalars becomes parametrically light, and eventually tachyonic, for  positive 
$\phi_I$.
The mass vanishes exactly at some finite value of $\f_I$, for which the  background geometry is still describing the dual of a confining field theory, in the presence of non-vanishing condensates. When this state is light, or tachyonic, the probe approximation does not capture it correctly, indicating that
the state has a non-trivial component along $h$, and hence is sourced by the trace 
of the stress-energy tensor, as expected by a dilaton.
We also notice that several of the heavy scalar states are not well captured by the probe approximation,
showing that mixing effects with the dilaton are not restricted to the lightest states.

\section{Free Energy}

To compute the free energy, we  write explicitly the boundary terms of the theory in $D=7$ dimensions:
\beqs
{\cal S}={\cal S}_7
\label{Eq:ActionD7Complete}
+\sum_{i=1,2}(-)^i\int\di^4x\di \zeta\di \eta \sqrt{-\tilde{\hat{g}}}\left[
\frac{K}{2}+\lambda_i
\right]_{\r=\r_i},
\eeqs
where $\tilde{\hat{g}}$ denotes the determinant of the pullback of the induced metric, $K$ is the Gibbons-Hawking-York (GHY) term
and $\lambda_i$ are  localised boundary potentials.

The potential terms are chosen according to the same prescription as in Ref.~\cite{Elander:2020ial}: in the UV
we replace $\lambda_2={\cal W}_2$, which allows one 
to cancel all the divergences and perform the programme of holographic 
renormalisation~\cite{Bianchi:2001kw,Skenderis:2002wp,Papadimitriou:2004ap}, while 
in the IR we impose $\lambda_1=-\frac{3}{2}\partial_{\rho}A(\rho)$, in such a way that the variational 
problem be well defined in the
presence of the IR boundary at $\r=\r_1$. 
The free energy density ${\cal F}$ is defined in terms of the complete on-shell action to be
\beqs
\int \di^4 x \di\zeta\di\eta {\cal F} &\equiv&-\lim_{\rho_2\rightarrow \infty}\lim_{\r_1\rightarrow \r_o}\,{\cal S}_{\rm on-shell}\,.
\eeqs
By making use of the equations of motion  we arrive at
\beqs
{\cal F}&=&
-\lim_{\rho_2\rightarrow \infty}\left.e^{4A-\chi}\left(
\frac{3}{2}\partial_{\rho}A+{\cal W}_2
\right)\right|_{\r_2}\,,
\label{Eq:free}
\eeqs
which is identical to Eq.~(5.22) of Ref.~\cite{Elander:2020ial}.

We  make use of the UV expansions of the background solutions of interest.
By replacing the  UV expansions in Eqs.~(\ref{Eq:UV1})-(\ref{Eq:UV4}) into
the form of the free energy density in Eq.~(\ref{Eq:free}), supplemented by the specific form 
of the superpotential ${\cal W}_2$ in Eq.~(\ref{Eq:W2}), we arrive at the expression:
\begin{widetext}
\beqs
{\cal F}&=&-\frac{e^{4A_U-\chi_U}}{120}\left(20 \phi_2\phi_4-135\chi_6+60\omega_6+12\sqrt{5}\phi_2^3\left(-2+\frac{3}{2}\log\left(\frac{\phi_2^2}{\kappa}\right)\right)\right)\,.
\label{Eq:freeUV}
\eeqs
\end{widetext}
 The divergence of the contribution to the free energy proportional to $\phi_2^2$ is cancelled by ${\cal W}_2$.
This  implies that,
 as for the circle reduction of the Romans supergravity~\cite{Elander:2020ial}, 
 the concavity theorems do not apply to ${\cal F}$.
This expression still contains a residual scheme-dependence, in the logarithmic
 term. We set  $\kappa= e^{-4/3}$,
 and hence our final expression for the free energy density is
\begin{align}
{\cal F}\,\, = \,\, -\frac{e^{4A_U-\chi_U}}{120}\bigg(20 \phi_2\phi_4&-135\chi_6+60\omega_6+\nonumber\\
&+18\sqrt{5}\phi_{2}^{3}\log(\phi_{2}^{2})\bigg)\,,
\label{Eq:freeUVfinal}
\end{align}
We also remind the reader that $\chi_6=0$ in the  background solutions of interest.

\begin{figure}[t]
\begin{center}
	\begin{picture}(220,160)
	\put(-15,0){\includegraphics[width=9cm]{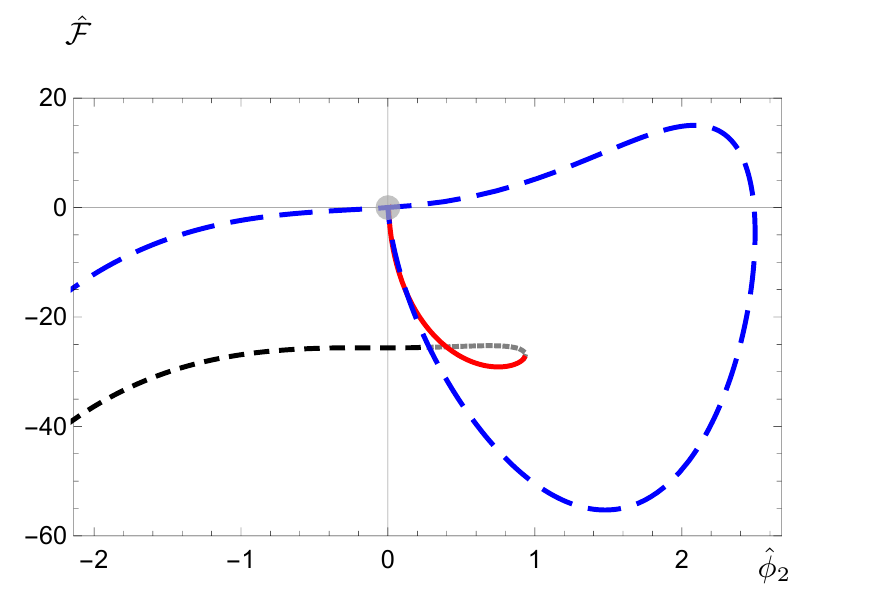}}
	\end{picture}
\caption{The free energy density $\hat{\cal F}={\cal F}\Lambda^{-6}$ as a function of 
the deformation parameter $\hat\phi_2=\phi_2\Lambda^{-2}$. The black (dashed), grey (short dashed) and red (solid) lines 
represent the stable, metastable and unstable portions  of the confining branch of solutions.
In blue (long dashed) we show the singular DW solutions. 
The susy solutions are represented by the grey disk at the 
origin of the plot.
}
\label{Fig:FreeEnergy}
\end{center}
\end{figure}

In Fig.~\ref{Fig:FreeEnergy} we show the free energy of the three classes of solutions,
as a function of the deforming parameter $\hat\phi_2\equiv \phi_2\Lambda^{-2}$,
and setting $A_U=0=\chi_U$. The parameters $\phi_4$ and $\omega_6$ are response functions, themselves determined non-linearly, on each branch of solutions, by the choice of $\phi_2$.
The SUSY solutions have ${\cal F}=0$.

The figure shows  evidence of the existence of a first-order phase transition.
The confining solutions minimise $\hat{\cal F}$ for negative $\phi_I$. For $\phi_I>\phi_I^c$,
with $\phi_I^c\simeq 0.039$ the critical value (corresponding to $\hat{\phi}_2^c\simeq 0.281$),
 the  singular  DW solutions have lower, finite  free energy density $\hat{\cal F}$, so that the solutions along the confining
branch are at best metastable when $\phi_I>\phi_I^c$, and eventually become unstable,
with one of their fluctuations becoming tachyonic when $\phi_I\gsim 0.447$.
Most interestingly, along the metastable branch, the lightest state becomes parametrically light,
before becoming tachyonic (see Fig.~\ref{Fig:SpectrumProbe}). 
The probe approximation fails to capture correctly its mass squared
when it is either small or negative. This eigenstate of the system is hence
an admixture containing a significant contribution from 
the trace of the fluctuation of the metric---we interpret this finding as evidence that the state is
approximately a dilaton.

\section{Outlook}

We presented evidence of the emergence of a parametrically light dilatonic state along the  
metastable portion of a branch of regular backgrounds
 of the supergravity system in $D=7$ dimensions
that yields also the Witten model, the first known
 holographic description of a four-dimensional confining theory~\cite{Witten:1998zw}.
Furthermore, the results of our analysis confirm, in the  rigorous  context of top-down holography,
the expectations from Ref.~\cite{Pomarol:2019aae} that along the stable portion of the regular branch
 a dilatonic state persists, but it is not parametrically light.
 
The metastable vacua, and the accompanying  parametrically light
 dilatonic state, are new findings. Comparison with Ref.~\cite{Elander:2020ial} indicates
 that this is a generic feature, which emerges in a broad class of theories.
It would be interesting to discover examples in which the phase transition is weaker, 
and the spectrum along the stable
branch  exhibits a light approximate dilaton.
 It would also be useful to identify the requirements a supergravity theory must 
fulfil for such features to emerge.

\vspace{0.0cm}
\begin{acknowledgments}
We thank Alex Pomarol for useful discussions.
The work of MP is supported in part by the STFC Consolidated Grants
 ST/P00055X/1 and ST/T000813/1. MP has received funding from the European Research Council (ERC) under the
European Union's Horizon 2020 research and innovation programme, grant agreement No 813942. 
JR is supported by STFC, through the studentship ST/R505158/1. 
\end{acknowledgments}



\end{document}